\documentclass[twocolumn,pre,floatfix,superscriptaddress,longbibliography]{revtex4}
\usepackage{color}
\usepackage{tabularx}
\usepackage{makecell}
\usepackage{dblfloatfix}
\usepackage{fixltx2e}
\usepackage{epsfig}
\usepackage{amsmath}
\usepackage{soul}
\usepackage{amssymb}
\usepackage{bm}
\usepackage{graphicx}
\usepackage{multirow}
\usepackage{wasysym}
\usepackage{varwidth}
\usepackage{enumitem}
\usepackage{algorithm, algpseudocode}
\usepackage{epsfig}
\usepackage{lipsum}
\usepackage{amsmath}
\usepackage[svgnames]{xcolor}
\usepackage{hyperref}
\usepackage{times}
\hypersetup{colorlinks=true,
            linkcolor=blue,
            citecolor=ForestGreen,
            urlcolor=DarkOrchid
}
\usepackage{etoolbox}
\makeatletter
\patchcmd{\math@cr@@@align}
  {\place@tag}
  {\bgroup\color{black}\place@tag\egroup}
  {}{}
\makeatother

\usepackage{soul}

\begin{document}

\title{Optimization and benchmarking of the thermal cycling algorithm}

\author{Amin Barzegar}
\affiliation{Department of Physics and Astronomy, Texas A\&M University,
College Station, Texas 77843-4242, USA}
\affiliation{Microsoft Quantum, Microsoft, Redmond, Washington 98052, USA}

\author{Anuj Kankani}
\affiliation{Department of Physics and Astronomy, Texas A\&M University,
College Station, Texas 77843-4242, USA}

\author{Salvatore Mandr{\` a}}
\affiliation{Quantum Artificial Intelligence Laboratory (QuAIL), NASA Ames Research Center, Moffett Field, California 94035, USA}
\affiliation{KBR, Inc., 601 Jefferson St., Houston, TX 77002, USA}

\author{Helmut G. Katzgraber}
\thanks{The work of H.~G.~K.~was performed before joining Amazon Web Services.}
\affiliation{Amazon Quantum Solutions Lab, Seattle, Washington 98170, USA}
\affiliation{AWS Intelligent and Advanced Compute Technologies, 
Professional Services, Seattle, Washington 98170, USA}
\affiliation{AWS Center for Quantum Computing, Pasadena, CA 91125, USA}

\begin{abstract}

Optimization plays a significant role in many areas of science and technology. Most of the industrial optimization problems have inordinately complex structures that render finding their global minima a daunting task. Therefore, designing heuristics that can efficiently solve such problems is of utmost importance. In this paper we benchmark and improve the thermal cycling algorithm [Phys.~Rev.~Lett.~79, 4297 (1997)] that is designed to overcome energy barriers in nonconvex optimization problems by temperature cycling of a pool of candidate solutions. We perform a comprehensive parameter tuning of the algorithm and demonstrate that it competes closely with other state-of-the-art algorithms such as parallel tempering with isoenergetic cluster moves, while overwhelmingly outperforming more simplistic heuristics such as simulated annealing.  
	
\end{abstract}

\date{\today}
\maketitle

\section{Introduction}

Optimization is ubiquitous in science and industry. From the search for the ground state of exotic states of matter such as high-temperature superconductors in physics \cite{stanev:18}, topology optimization in material science \cite{bendsoe:09}, lead optimization in pharmaceutical drug discovery \cite{keseru:06}, spacecraft trajectory optimization \cite{shirazi:18}, portfolio optimization in finance \cite{doering:16}, scheduling in transportation \cite{guihaire:08}, to speech recognition in artificial intelligence \cite{kinnunen:10}, to name a few.  One important category of optimization problems is combinatorial optimization, which is the search for the minima of an objective function within a finite but often large set of solutions. Paradigmatic examples are the traveling salesman problem, the geometrical packing problem \cite{karp:72}, graph coloring, the cutting stock problem, integer linear programming, etc. Many of these problems are NP-hard, in the sense that the worst-case time to find the optimum scales worse than a polynomial in the size of the input. Moreover, these problems are not only computationally hard to solve in the worst-case, but also in the typical case. These problems have a rough energy (cost function) landscape, consisting of numerous metastable states. Therefore, heuristics based on local search---e.g., the greedy algorithm \cite{cormen:90}---tend to perform poorly on these types of problems as they can easily become trapped in local minima \cite{bendall:06}. 

One way to circumvent this difficulty is to use a stochastic process such as Metropolis dynamics \cite{metropolis:49} to randomly access different parts of the phase space. An example of an algorithm that utilizes such random sampling is simulated annealing (SA) \cite{kirkpatrick:83}. Simulated annealing is a Markov Chain Monte Carlo (MCMC) process where a series of quasi-equilibrium states are visited by following an ``annealing schedule'' during which the system is gradually cooled from a sufficiently high temperature to a target low temperature. The goal is to guide the stochastic process through an occasionally complex energy landscape toward the low-lying states. At high temperatures, the random ``walker'' can take long strides across phase space, thus allowing for the exploration of configurations far away in Hamming distance. As the system is cooled, the exploration domain of the walker is reduced according to the Gibbs distribution, and it eventually lands in a low-lying state. There is no guarantee that this is the true optimum, unless an (impractical) infinite annealing time is used \cite{geman:84}. Because SA is stochastic in nature, running many such processes in parallel can increase the chance of finding the true optimum. Nevertheless, without establishing a way for the phase-space information gathered by the random walkers to be shared, mere replication of a simulated annealing process will not yield any meaningful speedup. Multiple Markov chain algorithms such as path integral Monte Carlo (PIMC) \cite{trotter:59,suzuki:71,suzuki:76,suzuki:93,landau:00,troyer:03}, parallel tempering (PT) \cite{geyer:91,hukushima:96}, and population annealing (PA) \cite{hukushima:03,machta:10,wang:15e,amey:18,barzegar:18a} take advantage of this ``collective knowledge'' to efficiently probe the solution space of a problem.  

Closely related to the genetic local search approaches \cite{brady:85,muhlenbein:88,freisleben:96,merz:98}, the thermal cycling algorithm (TCA) \cite{moebius:97,moebius:05} is another heuristic that integrates the power of parallel annealing processes with the utility of local search methods. The annealing part of this algorithm ensures that the phase space can be visited ergodically, whereas the local search part biases the dynamics toward the lower-energy states. When introduced almost twenty years ago, thermal cycling was shown to outperform simulated annealing in solving some limited instances of the traveling salesman problem. Despite the early indications that TCA might be a useful tool in dealing with hard optimization problems, it has not been carefully benchmarked and hence widely adopted by the optimization community. Here we reintroduce the thermal cycling algorithm and outline the basic pseudocode. In addition, we conduct a comprehensive parameter optimization of TCA using synthetic planted problems, where we compare the performance of TCA to a number of modern optimizers, including simulated quantum annealing (SQA). In order to quantify the efficiency of the aforementioned heuristics, we study how their time to solution (TTS) \cite{boixo:14,ronnow:14a} scales with the problem size. Our results show that when optimized properly, TCA can indeed be competitive with the state-of-the-art heuristics and therefore should be included in physics-inspired optimization platforms.

The paper is structured as follows. In Sec.~\ref{sec2:analysis} we explain the analysis techniques followed by the details of the algorithm in Sec.~\ref{sec3:TCA}. Section \ref{sec4:results} is dedicated to the benchmarking results of the study. Concluding remarks are presented in Sec.~\ref{sec5:conclusion}.

\section{Details of Analysis}
\label{sec2:analysis}

The cost function that we minimize in this benchmarking study is a 2-local Ising spin system, i.e.,
\begin{align}
\mathcal{H}=\sum_{i}^N\sum_{j\in\mathcal{N}_i}J_{ij}s_is_j +\sum_i^Nh_is_i,
\end{align}
where $N$ is the total number of variables, $\mathcal{N}_i$ is the adjacency list of the $i$'th lattice site, $J_{ij}$ is the coupling between spin $s_i$ and $s_j$, and finally $h_i$ is an external field applied to spin $s_i \in \{\pm 1\}$. 

Most algorithms involve multiple parameters that need to be carefully tuned to observe the true asymptotic scaling. As such, a comprehensive hyperparameter optimization is in order. For benchmarking, we use synthetic problems whose ground state is unique and known beforehand. Here, we use the deceptive cluster loop (DCL) problems \cite{mandra:18} that are specifically designed for testing the performance of the D-Wave \cite{dwave} quantum annealer against classical algorithms. DCL's are inspired by the original frustrated cluster loop (FCL) problems \cite{hen:15a,king:19}, which have a ferromagnetic planted ground state defined on a Chimera graph \cite{bunyk:14}. The Chimera topology consists of a two-dimensional lattice of fully-connected bipartite $K_{4,4}$ cells in which all qubits are coupled ferromagnetically. The entire $K_{4,4}$ unit cell can, therefore, be viewed as one virtual variable. The cells are then connected via randomly-chosen frustrated loops. The magnitude of inter-cell couplings are capped at a finite value $R$ that adds local ``ruggedness" to the problems \cite{king:15, king:17}. The hardness of the FCL instances can be tuned by varying the density of the frustrated loops, often denoted by parameter $\alpha$. In the DCL problems, the inter-cell couplers are multiplied by a scaling factor $\lambda$. Depending on the value of $\lambda$, the internal structure of the cells can be masked or accentuated, thus deceiving the annealers to spend more time optimizing the local structures rather than finding the global minimum.

As the measure of performance, we use the time to solution (TTS) \cite{boixo:14,ronnow:14a} that is defined in the following way:
\begin{align}
{\rm TTS}(\alpha)=n(\alpha)\,\tau_{\rm run},\label{tts}
\end{align} 
where $n(\alpha)$ is the number of times that the algorithm must be repeated, for a given parameter set $\alpha$, to find the ground state at least once with a desired probability of $p_{\rm d}$. $\tau_{\rm run}$ is the average run time, conventionally measured in microseconds. If we assume that the {\em success probability}, {\em i.e.}, the chance of hitting the ground state in a single run of the algorithm is $p_{\rm s}(\alpha)$, then one can show from the binomial distribution that 
\begin{align}
p_{\rm d} = \sum_{k\geq 1}^{n}{n\choose k}p_{\rm s}^k(1-p_{\rm s})^{n-k}=1-(1-p_{\rm s})^n.\label{binomial}
\end{align}
We may now use the above expression to find $n(\alpha)$ in Eq.~\eqref{tts}:
\begin{align}
n(\alpha)=\frac{\log[1-p_{\rm d}]}{\log[1-p_{\rm s}(\alpha)]}.
\label{repetitions}
\end{align}
It is customary to set the desired probability in Eq.~\eqref{repetitions} to a high confidence value of $p_{\rm d} = 0.99$. Because the TTS is a function of the algorithm parameters, a thorough optimization of the parameters must be performed to reliably compare heuristics based upon it. Note that the optimization is often multidimensional, which makes the benchmarking a relatively laborious task. For each set of parameters $\alpha$ and each problem instance, we repeat the runs $100$ times and calculate the success probability $p_{\rm s}(\alpha)$ as the percentage of the ground state hits. This process is repeated for all instances, in this case, $100$, to calculate the median TTS, and the error bars are estimated using the bootstrap method. The above procedure is carried out for many other parameter-set values, and the optimal parameters are identified as the global minimum point of the TTS function. Having calculated the optimal TTS for all problem sizes ($N = 8L^2$), we can study the scaling behavior of the algorithm, which is often an exponential, {\em i.e.}, 
\begin{align}
{\rm TTS}_{\rm opt}\sim 10^{a + bL}.\label{scaling}
\end{align}
The scaling exponent $b$ determines the performance of an algorithm in the asymptotic limit, whereas $a$ is a constant offset that depends on the factors nonintrinsic to the algorithm, such as hardware speed, code efficiency, etc. Therefore, a relatively unbiased way to compare different algorithms is to focus on the scaling exponent.

\section{Thermal Cycling Algorithm}
\label{sec3:TCA}

The thermal cycling algorithm works by periodically heating and cooling an ensemble of states while following a decreasing temperature schedule. The ensemble is prepared by selecting $N_{\rm p}$ lowest energy states among $N_0$ quenched random configurations. Starting from the initial inverse temperature $\beta_{\rm i}=0$, the above pool of states is annealed toward a final inverse temperature of $\beta_{\rm f}$ in $N_{\rm T}$ steps. At a given temperature, some energy is deposited into the ensemble states using $N_{\rm s}$ Metropolis updates (heating) followed by an immediate quench via a local search method (cooling). If any of the resulting states are lower in energy than the original set, they are replaced in the pool. The heating-cooling cycle is repeated $N_{\rm c}$ times at a fixed temperature. In practice, the above process steers the ensemble toward the low-lying states while ensuring that metastable configurations do not hinder the dynamics. The temperature is then reduced, and the cycles start over. In Algorithm~\ref{algorithm:TCA}, we present a concise outline of the thermal cycling algorithm.

\begin{algorithm}[t!]
  \caption{Thermal Cycling}
   \begin{algorithmic}[1]
   \State Randomly initialize $N_0$ configurations of the problem. 
   \State Quench each of the $N_0$ states using a local search algorithm. 
   \State Construct a pool of states by selecting $N_{\rm p}$ states with the lowest energy from the above quenched states. 
	\State Build a list of lattice sites by comparing the spins on a given site between all pools states. If all aligned, add the site to the list. 
   \For { $N_{\rm T}$ steps starting from $\beta=0$ until $\beta=\beta_{\rm f}$}
		\For {$N_{\rm c}$ cycles}   
			\State Pick a random state from the pool.
			\State {
				\begin{varwidth}[t]{\linewidth}
					Add heat to the pool state using $N_{\rm s}$ Metropolis sweeps \par
		        	\hspace{-2cm} at $\beta$, excluding the spins in the site list. 
			  	\end{varwidth}
			 }
			\State Quench the selected state.
			\If {lower energy is achieved}
				\State Replace the old state in the pool with the new one. 
				\State{
					\begin{varwidth}[t]{\linewidth}
						Rebuild the site list by comparing the spins between \par 
						\hspace{-3.6cm} the updated pool states. 
				   \end{varwidth}
				}
			\EndIf
		\EndFor
   		\State {
   			\begin{varwidth}[t]{\linewidth}
   				 Increase $\beta\rightarrow \beta+\Delta \beta$ in which the step size  $\Delta\beta$  is \par 
   				    \hspace{-1.9cm} usually constant, i.e, linear schedule.
   			\end{varwidth}
   		}
   \EndFor
   \State Identify the pool state with the lowest energy as the solution of the problem. 
   \end{algorithmic}
   \label{algorithm:TCA}
\end{algorithm}

The main advantage of thermal cycling is the possibility of using a variety of variable-update classes in the quenching phase. By using more complex updates, exponentially many smaller local minima can be skipped in favor of lower-energy and configurationally more differing ones. This, however, does not necessarily translate to increased efficiency of the algorithm as the implementation overhead associated with those complex moves can negate the overall gain. Thus, there must be a trade-off between the complexity of the moves and the speedup owing to the reduced metastability. One of the simplest updates is a single-spin greedy move (SSGM) in which the most unstable spins ({\em i.e.,} spins with the largest positive local fields) are flipped in a sequential fashion until no further improvement can be made. Another subset of the move classes are the double-spin random moves (DSRM), which consist of first attempting to flip a randomly-chosen spin by itself, and if this is rejected, then trying to flip it together with one of the neighboring spins (looked up sequentially) that results in lowering the overall energy. The updates stop when the rejections accumulate to the total number of bonds in the problem.  Another important type of move that we have studied here is the Lin-Kernighan cluster move (LKCM) that is based on the famous Lin-Kernighan algorithm \cite{lin:73,helsgaun:00} which is considered, to date, one of the most efficient heuristics for solving the traveling salesman problem (TSP).  In a LKCM, a cluster of spins of size $M$ is constructed starting from the most unstable spin and then appending the neighboring spins to it until the total cost of flipping the cluster becomes positive. The LKCM is essentially a $k$-opt local search algorithm \cite{flood:56,croes:58,lin:65}, where $k$ is determined from a sequence of partial costs, {\em i.e.}, $\{c_1,c_2,\ldots, c_M\}$. Here, $c_1$ is the cost of flipping the first spin, $c_2$ is the cost of flipping the first and the second spin together, and so on. One then flips the set of $k$ spins with the lowest partial cost, $c_k$. We report the performance of each of the above move classes in the next section. 

Another important point that one has to bear in mind is the duration of the heating phase. As mentioned earlier, the heating part of TCA ensures that the algorithm remains dynamic in spite of being quenched to often deep local minima. This requires that the system is subjected to a sufficient number of Metropolis updates. On the other hand, to preserve the gains of the previous cycles, the equilibration must be terminated in early stages. Otherwise, the system might end up in a configuration too far away in the phase space.

\begin{figure}[t!]
\begin{center}
\includegraphics[width=\columnwidth]{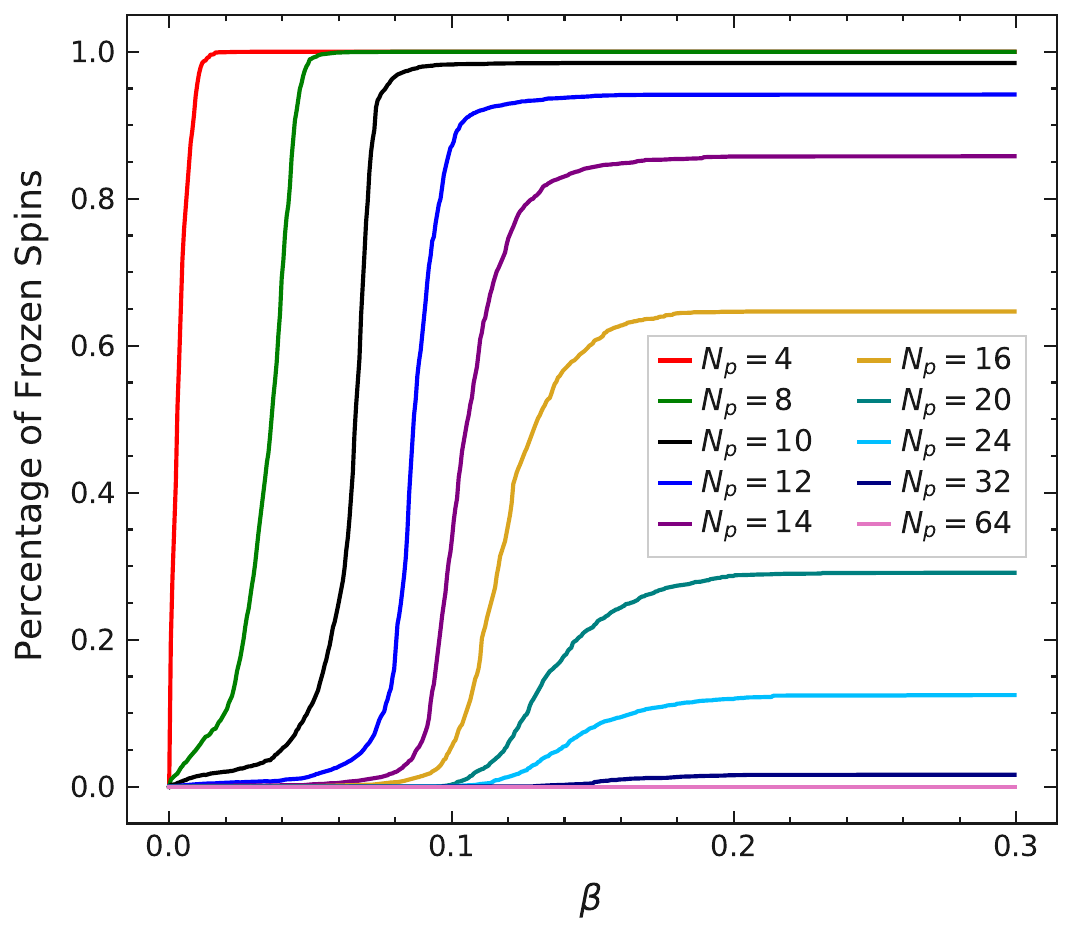}
\caption{
The percentage of frozen spins in TCA for the system size $L\!=\!12$ versus inverse temperature $\beta$. The colors on the lines represent different pool sizes $N_{\rm p}$. With only a few walkers, there is more chance of agreement between the spin values at a given site, hence the system collapses to a single walker early on during the annealing process, whereas with too many walkers, the freezing mechanism ceases to function. 
}
\label{fig1:frozen_percentage}
\end{center}
\end{figure}

During the cycling process, the states in the pool are treated independently from one another. This has the potential pitfall that some of the states in the pool might wander off to energetically unfavorable parts of the configuration space. As mentioned earlier, this shortcoming can be alleviated by establishing an interaction between the pool states. One way to do this is to freeze the variables that are common among all of the states. We can justify this reduction by realizing that if the states in the pool have a feature in common, it is very likely that the feature will also appear in the ground-state configuration. Note that this step is closely related to metaheuristics such as tabu search \cite{glover:89,glover:90} as well as self-avoiding random dynamics on integer complex systems (SARDONICS) \cite{hamze:11}, search for backbones \cite{schneider:96,zhang:04,wang:11b, wang:13d} often used in genetic type algorithms, and sample persistence \cite{chardaire:95,karimi:17a} which has been used in conjunction with algorithms such as simulated annealing as well as simulated quantum annealing. 
\pagebreak
\onecolumngrid
\begin{table*}[t!]
\begin{center}
\caption{
Optimal parameters of the studied algorithms for different linear problem sizes $L$. Here, $N_{\rm T}$ represents the total number of algorithmic steps in each heuristic. For example, in SQA, one algorithmic step involves building a Wolff cluster \cite{wolff:89} in the imaginary time direction by sweeping randomly through all lattice sites, whereas in PT, an algorithmic step in defined as a Metropolis sweep over each replica followed by a tempering exchange move. $\beta_{\rm i}$ and $\beta_{\rm f}$ are the highest and lowest temperatures that the algorithm operates between. TCA involves the additional parameters of $N_{\rm p}$, $N_{\rm c}$, and $N_{\rm s}$ which are the pool size, the number of heating-cooling cycles per temperature, and the number of Metropolis sweeps, respectively. In SQA, $N_{\rm p}$ determines the number of Suzuki-Trotter slices while $\beta_{\rm c}=\beta_{\rm q}/N_{\rm p}$ sets the ``classical" inverse temperature with $\beta_{\rm q}$ being the temperature at which the underlying quantum annealing process is performed. Finally, for PT $M$  is the number of temperatures which are spaced as a geometric sequence. When PT is accompanied with ICM updates, $2M$ replicas are used.}\label{table1:optimal_parameters}
\begin{tabular*}{\textwidth}{@{\extracolsep{\fill}} p{1cm} l l p{0.5cm} l l p{0.5cm} l l l l l p{0.5cm} l l l p{0.5cm} l l l l r}
 \hline
 \hline
  & \multicolumn{2}{c}{SA} & & \multicolumn{2}{c}{SA+DSRM} & & \multicolumn{5}{c}{TCA} & & \multicolumn{3}{c}{SQA} & & \multicolumn{4}{c}{PT}\\
 \hline
 $L$ & $N_{\rm T}$ & $\beta_{\rm f}$ &  & $N_{\rm T}$ & $\beta_{\rm f}$ & & $N_{\rm T}$ & $N_{\rm p}$ & $N_{\rm c}$ & $N_{\rm s}$ & $\beta_{\rm f}$ & & $N_{\rm T}$ & $N_{\rm p}$ & $\beta_{\rm c}$ & & $N_{\rm T}$ & $M$ & $\beta_{\rm i}$ & $\beta_{\rm f}$ \\
 \hline
 $8$ & $2896$  & $0.25$ & & $3242$  & $0.25$  & & $128$ & $16$ & $32$ & $16$  & $0.20$ & & $65536$ & $32$ & $1.0$ & & $2220$ & $8$ & $0.10$ & $0.20$\\
$9$ & $8192$ & $0.25$ & & $7529$  & $0.25$ & & $256$  & $16$ & $32$ & $16$ & $0.25$ & & $92681$ & $32$ & $1.0$ & & $3246$ & $8$ & $0.10$ & $0.25$\\
$10$ & $11585$  & $0.30$ & & $11621$  & $0.30$ & & $512$  & $16$ & $64$ & $32$  & $0.25$ & & $185363$ & $32$ & $1.0$ & & $5686$ & $16$ & $0.10$ & $0.25$\\
$11$ & $46340$  & $0.30$ & & $26989$  & $0.30$ & & $512$  & $16$ & $128$ & $32$  & $0.30$ & & $262144$ & $32$ & $1.0$ & & $10787$ & $16$ & $0.10$ & $0.30$\\
$12$ & $65536$  & $0.30$ & & $41127$  & $0.30$ & & $724$  & $16$ & $128$ & $32$ & $0.30 $ & &  $262144$ & $32$ & $1.0$ & & $17695$ & $16$ & $0.10$ & $0.30$\\
$13$ & $92680$  & $0.30$ & & $62675$  & $0.30$ & & $724$  & $16$ & $128$ & $32$ & $0.30$ & & $370727$ & $32$ & $1.0$ &  & $21899$ & $24$ & $0.10$ & $0.30$\\
$14$ & $262144$ & $0.30$ & & $95514$  & $0.30$ & & $1024$  & $16$ & $128$ & $32$ & $0.30$ & & $524289$ & $32$ & $1.0$ & & $34954$ & $24$ & $0.10$ & $0.30$\\
$15$ & $1048576$  & $0.30$ & & $145556$  & $0.30$ & &  $1024$  & $16$ & $128$ & $32$ & $0.30$ & & $741455$ & $64$ & $1.0$ & & $58909$ & $32$ & $0.10$ & $0.30$\\
$16$ & $1482910$  & $0.30$ & & $221815$  & $0.30$ & & $1448$  & $16$ & $128$ & $32$ & $0.30$ & &  $1048576$ & $64$ & $1.0$ & & $96183$ & $32$ & $0.10$ & $0.30$\\
\hline
\hline
\end{tabular*}
\end{center}
\end{table*}
\twocolumngrid
\noindent

In Fig.~\ref{fig1:frozen_percentage}, we show the percentage of frozen spins versus the inverse temperature $\beta$ for the system size $L=12$ in which different colors represent various pool sizes $N_{\rm p}$. It is interesting to observe that the freezing mechanism is only helpful when there is a moderate number of walkers. In other words, with too few walkers the entire ensemble collapses to one state very early on in the annealing schedule, whereas with too many walkers, the probability of all the pool states agreeing on the value of a particular spin becomes exceedingly low, hence rendering the freezing practically irrelevant. On the other hand, with population size around $N_{\rm p}=16$, the percentage saturates at an optimal value of roughly $60\%$ such that a considerable number of degrees of freedom are preserved allowing for independent random walks to continue while still restraining the walkers from spreading to far from one another in the configurational space.  

\section{Results}
\label{sec4:results}
In this study, we compare TCA to simulated annealing (SA) \cite{kirkpatrick:83}, simulated quantum annealing (SQA) \cite{suzuki:76}, parallel tempering (PT) \cite{geyer:91,hukushima:96}, and parallel tempering with isoenergetic cluster moves (PT+ICM) \cite{zhu:15b}. SQA is the classical implementation of the quantum annealing process \cite{finnila:94, kadowaki:98} in which the system is initialized in the ground state of a simple Hamiltonian and adiabatically \cite{born:28} deformed into a target Hamiltonian whose ground state is difficult to find. PT is a Monte Carlo algorithm that efficiently samples the equilibrium configurations of a system using the replica-exchange technique. The ICM update---which consist of rearranging a large collection of variables by inspecting the overlap between two replicas---is extremely effective for low connectivity graphs in which the cluster percolation threshold is small.
For this benchmarking study we have generated $100$ DCL instances for each linear size ranging from $L=8$ to $L=16$. The DCL parameters are fixed to a relatively hard regime of $\alpha=0.24$, $R=1$, and $\lambda=3.0$ \cite{mandra:18}. For each studied algorithm, we optimize the parameters via a grid search within its parameter space. All of the simulations are done on a single thread using Intel Xeon E5-2680 v4 2.40GHz  and Intel Xeon E5-2673 v4 2.30GHz processors. In Table \ref{table1:optimal_parameters}, we have listed the optimal parameters of the thermal cycling algorithm as well as the other studied heuristics.

 Note that since we are dealing with a high-dimensional optimization space, global optimality is neither guaranteed nor necessarily unique. We observe that most of the TCA parameters are robust with respect to the problem size and the number of annealing steps $N_{\rm T}$---much like SA---is the only varying parameter. This is valuable information as it eliminates the necessity of a full parameter optimization in a practical implementation of the algorithm. In reality, the total effort in a TCA simulation is roughly proportional to $N_{\rm T}N_{\rm p}N_{\rm c}N_{\rm s}$ with some additional overhead caused by the local search. This suggests that correlation between the above parameters can be expected. For instance, similar performance can be achieved by increasing the number of walkers or by extending the annealing schedule while having a moderate pool size. 

\begin{figure}[t!]
\begin{center}
\includegraphics[width=\columnwidth]{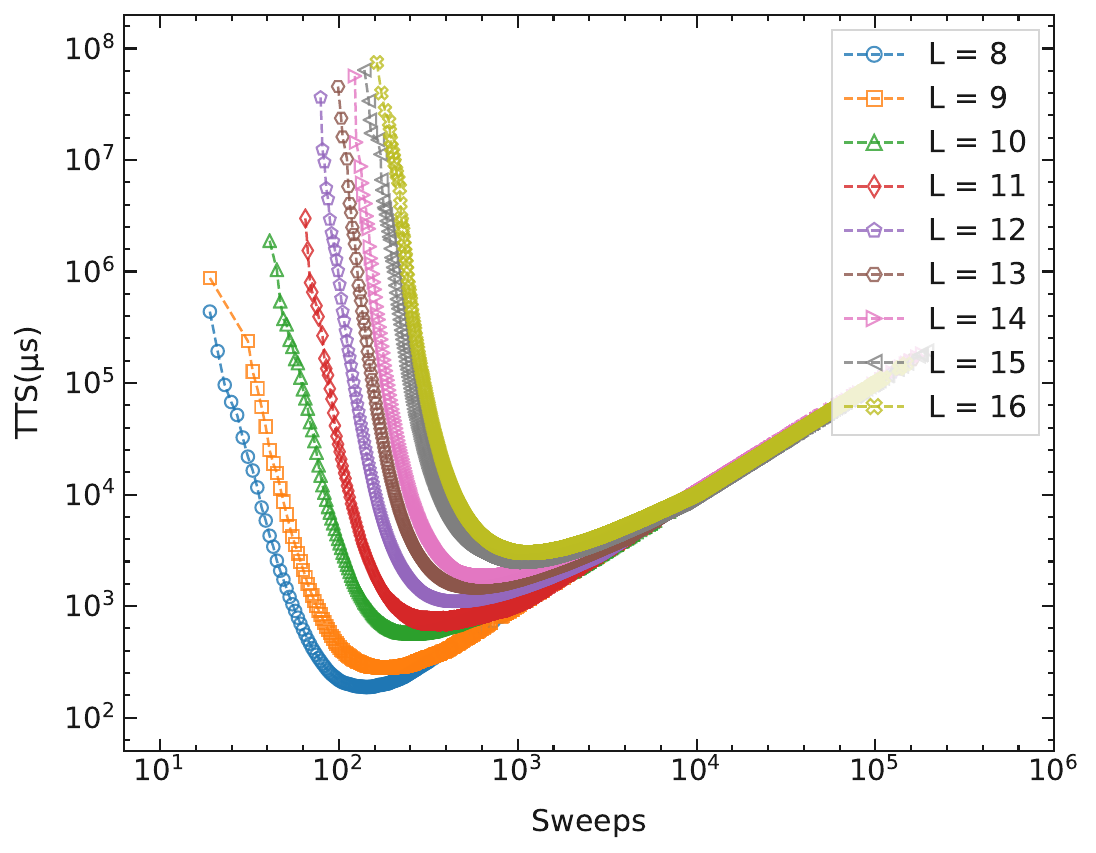}
\caption{
Optimization of total sweeps for PT+ICM using various system sizes. The minimum TTS points correspond to the optimal sweep values. 
}
\label{fig1:sweep_optimization_PT+ICM}
\end{center}
\end{figure}

\begin{figure}[t!]
\begin{center}
\includegraphics[width=\columnwidth]{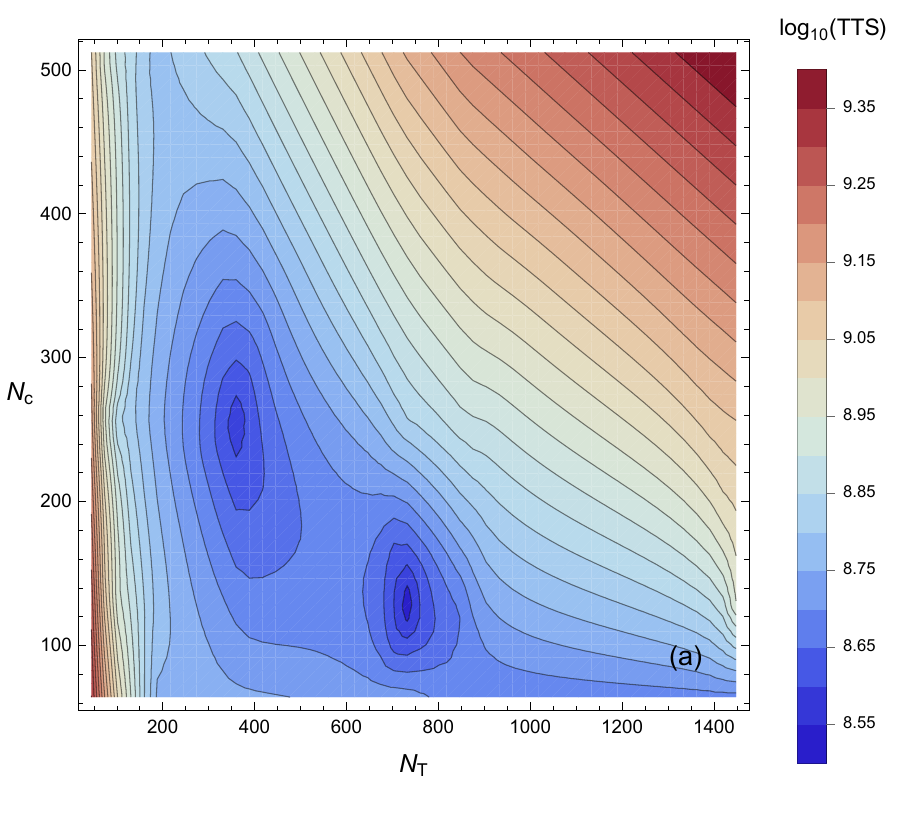}
\hspace*{-1cm}\includegraphics[width=0.96\columnwidth]{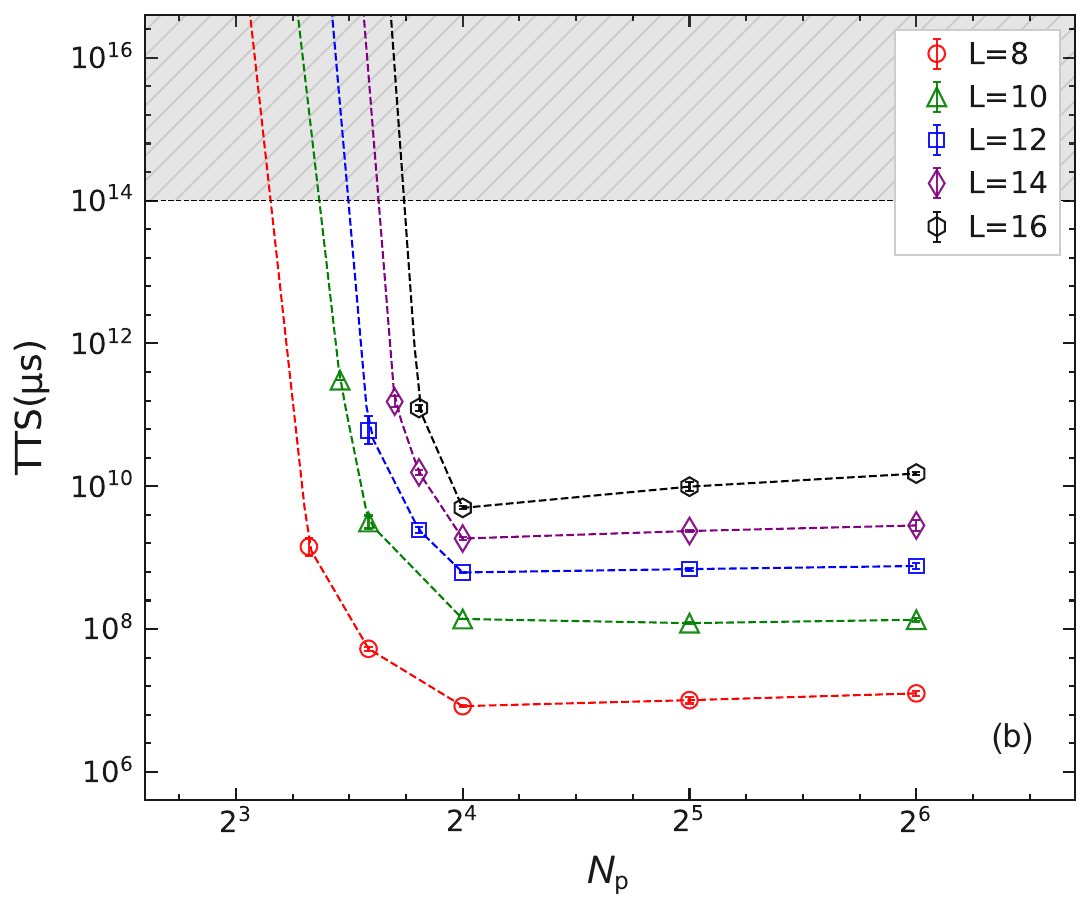}
\caption{
(a) Parameter tuning of the thermal cycling algorithm using deceptive cluster loop (DCL) problems of linear size $L\!=\!12$. The figure shows the number of cycles $N_{\rm c}$  versus the number of annealing steps $N_{\rm T}$, with the rest of the parameters fixed to the values listed in Table \ref{table1:optimal_parameters}. The color map shows the TTS values. Two minima with similar depths are observed. The lowest minimum corresponds to the optimal parameters. (b) Optimization of the pool size $N_{\rm p}$ of TCA for a various system sizes. Above the horizontal dashed line, none of the problems can be solved in 100 attempts. Across all problem sizes studied $N_{\rm p}=16$ is sufficient.}
\label{fig3:TCA_parameter_optimization}
\end{center}
\end{figure}

In Figures \ref{fig1:sweep_optimization_PT+ICM} and \ref{fig3:TCA_parameter_optimization} we show some examples of such parameter optimization. Figure \ref{fig1:sweep_optimization_PT+ICM} shows time to solution versus the total number of sweeps in PT+ICM for various problem sizes. The minimum of the curve marks the optimal sweep values. Figure \ref{fig3:TCA_parameter_optimization}(a) illustrates a two-dimensional cross section of the parameter space of TCA for the system size $L=12$. The rest of the parameters are fixed to the values listed in Table \ref{table1:optimal_parameters}. Here, the color map represents the TTS values in a logarithmic scale. The axes show the number of cycles $N_{\rm c}$ and the number of annealing steps $N_{\rm T}$. We observe two minima with comparable depth within error bars, corroborating the fact that $N_{\rm T}$ and $N_{\rm c}$ are anti-correlated. Figure \ref{fig3:TCA_parameter_optimization}(b) shows the TTS as a function of the pool size $N_{\rm p}$. Beyond the horizontal dashed line marked by the hatched region, none of the benchmark problems can be solved in $100$ independent attempts. It is interesting to observe that having a sufficiently large number of walkers is essential for the efficiency of the algorithm. Although it is intuitive that having more walkers will increase the chance of finding the ground state, the additional computational effort of doing so negates any potential gains as seen from the flat regions in the TTS curves. 

\begin{figure}[t!]
\begin{center}
\includegraphics[width=\columnwidth]{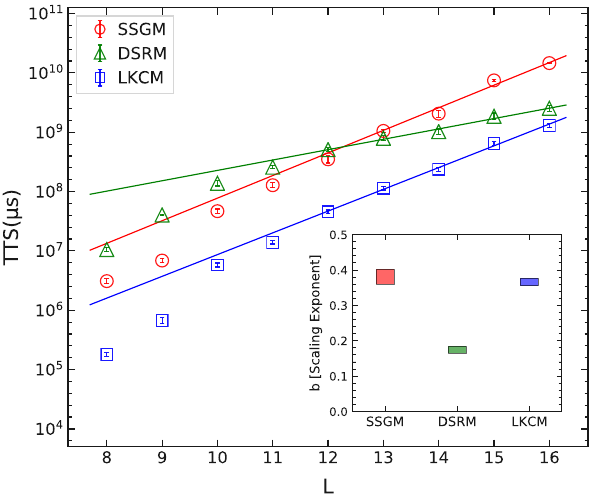}
\caption{
Main Panel: Time to solution (TTS) of the thermal cycling algorithm versus the linear problem size $L$. Various quenching schemes consisting of single-spin greedy moves (SSGM), double-spin random moves (DSRM), and Lin-Kernighan cluster moves (LKCM) are displayed. The lines represent a linear fit in the logarithmic (base 10) scale to the five largest system sizes. We observe a sizable constant speedup with the LKCM although it diminishes at larger system sizes. Inset: Scaling exponent $b$ corresponding to the slope of the linear fits in the main panel. The height of the boxes represents the error bars. The best scaling is obtained when TCA is used in conjunction with DSRM. These moves are simple enough to cause minimal overhead, yet complex enough to considerably reduce the number of the metastable states. 
}
\label{fig4:scaling_TCA}
\end{center}
\end{figure}

In Fig.~\ref{fig4:scaling_TCA} we show the scaling results of the thermal cycling algorithm with various move classes--that is, SSGM, DSRM, and LKCM, as explained in Sec.~\ref{sec3:TCA}. The main panel of Fig.~\ref{fig4:scaling_TCA} displays the optimal TTS values for different system sizes $L$, with the lines fitted to the largest five system sizes. The inset of the figure shows the scaling exponent $b$ in Eq.~\eqref{scaling} obtained from the linear fits. Note that the height of the boxes represents the error bars. It is clear that the double-spin random moves (DSRM) are significantly more efficient than the other two types of move classes that we have studied here. It is interesting to note that the Lin-Kernighan cluster moves (LKCM) become less efficient as the problem size increases despite giving almost two orders of magnitude in constant speedup for smaller systems sizes. We speculate that this is due to the increased overhead of constructing a long sequence of partial costs---which requires building a large cluster of spins as we explained in Sec.~\ref{sec3:TCA}---relative to the optimal subcluster that is flipped in the end. It is worth mentioning here that the efficiency of the LKCM is to some extent topology-dependent because the updates are, in essence,  cluster moves, and therefore might perform better when implemented on a different set of problems.

\begin{figure}[t!]
\begin{center}
\includegraphics[width=\columnwidth]{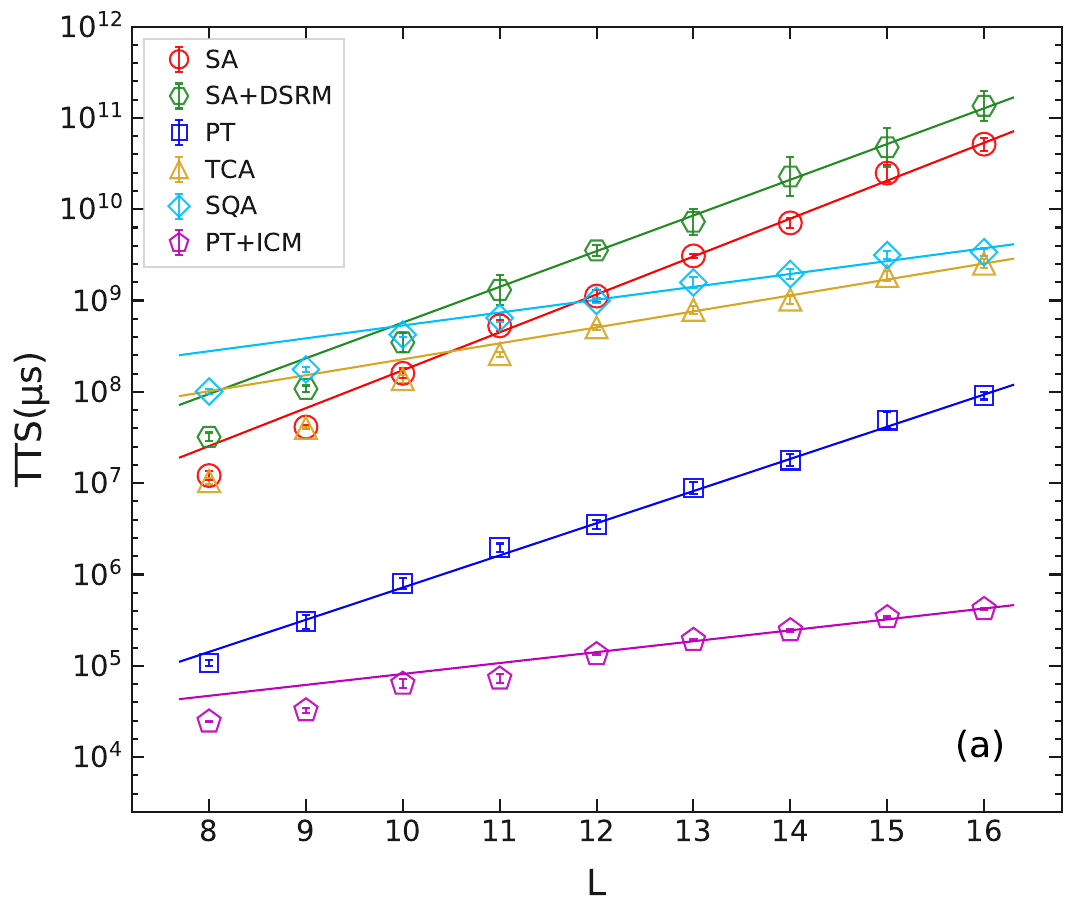}
\includegraphics[width=\columnwidth]{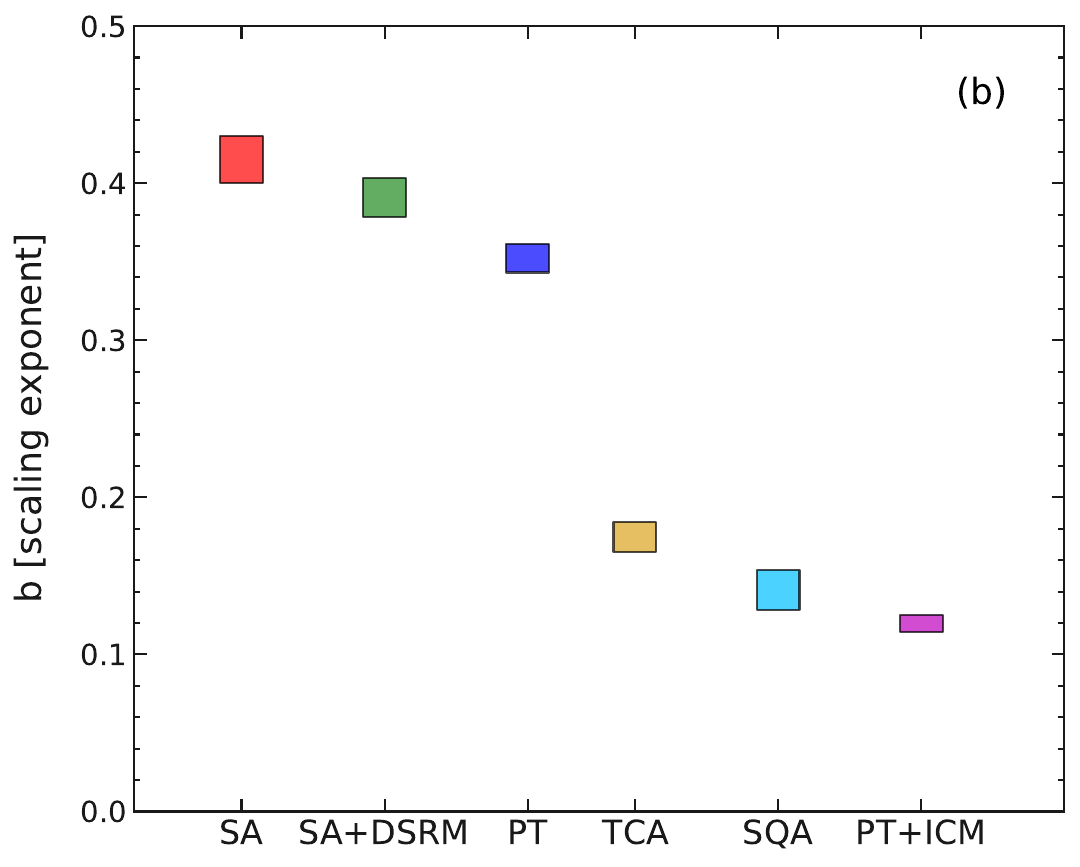}
\caption{
Comparison between the scaling results of the studied algorithms using the deceptive cluster loop (DCL) problems. (a) Time to solution (TTS) versus the linear problem size $L$ for various algorithms. Note that TTS is given on a logarithmic (base 10) scale. (b) Scaling exponent $b$ in Eq.~\eqref{scaling} for various algorithms. The height of the boxes represents the error bars. The TCA data points correspond to the best performing quenching scheme, {\em i.e.}, DSRM, shown in Fig.~\ref{fig4:scaling_TCA}. TCA scales better than SA and PT, and even comparable to SQA within the error bars. PT+ICM shows the best scaling. However, the latter is due to the great efficiency of the ICM updates in sampling the DCL phase space which is non-representative for denser industrial problems. Unlike in TCA, the addition of DSRM quenches to SA (SA+DSRM) does not improve the performance.
}
\label{fig5:scaling_all_solvers}
\end{center}
\end{figure}

In Fig.~\ref{fig5:scaling_all_solvers}(a), we show the scaling curves of various algorithms, corresponding to the optimal values of their parameters, as a function of the linear size $L$.  As before, the TTS values are reported in the logarithmic (base 10) scale where the linear fits are interpreted as exponential scalings. There is a considerable constant offset associated with PT and PT+ICM, which is due to the use of a highly-optimized implementation by Salvatore Mandr{\`a} as a part of NASA/TAMU Unified Framework for Optimization (UFO). Figure \ref{fig5:scaling_all_solvers}(b) shows the scaling exponent $b$ [the slope of the linear fit in  Fig.~\ref{fig5:scaling_all_solvers}(a)]. We observe that TCA (with DSRM) scales overwhelmingly better than SA, in agreement with previous TSP studies. It is also more efficient than PT and even competitively close to SQA. Note that PT is already established as a powerful heuristic in many optimization-related applications. SQA and PT+ICM show the best performances among the studied solvers. This can be ascribed to the structure of the DCL problems that involve tall but thin barriers that can be easily tunneled through using SQA. Isoenergetic cluster moves are also well-suited for the DCL problems as they cause large rearrangements of the variables, resulting in an efficient sampling of the configuration space. 

We have also included SA+DSRM in which simulated annealing is augmented by the double-spin random moves used in TCA as explained earlier in Sec.~\ref{sec3:TCA}. We notice that unlike TCA, the improvement is negligible which highlights the fact that the efficiency of TCA does not solely arise from the use of complex quench moves but rather because of the population-based nature of the algorithm. With a single random walker in SA, the zero-temperature quenches tend to strongly disturb the equilibrium distribution of states obtained by the finite-temperature Metropolis moves. On the contrary, with multiple walkers, the effect of such quenches is propagated slowly through the population allowing the downhill moves to be exploratory rather than disruptive. This can be best seen in Fig.~\ref{fig3:TCA_parameter_optimization}(b) where TTS values increase dramatically once the pool size drops down to a handful of walkers.

\section{Conclusion}
\label{sec5:conclusion}

In this paper we have thoroughly benchmarked the thermal cycling algorithm. Our results demonstrate that TCA is a competitive heuristic for solving problems with complex structures as it takes advantage of repeated heating and cooling to push the system toward the lower-energy states while ensuring that the system does not get trapped in an excited state. By reducing the variables among the TCA replicas, the stochastic process can be further accelerated, and the system can be guided more effectively toward the global minimum using the collective ``memory" of the solution pool. Having carefully tuned the parameters, we show that TCA can be as effective as the state-of-the-art algorithms such as SQA, while overpowering SA and PT by great margins in the asymptotic scaling. 

Due to the special structure of the DCL problems, which involve tall yet narrow barriers, SQA and PT+ICM outperform TCA because they utilize quantum effects and cluster updates to bypass those barriers. The true advantage of TCA might be revealed when using dense graphs with broad barriers, where PT+ICM and SQA would naturally struggle. TCA also lends itself to being integrated with ICM updates because it involves simultaneous annealing of many system replicas. This is in close analogy with to the Iterative Partial Transcription (IPT) algorithm \cite{moebius:99a, moebius:99b}.  It has been shown by Ochoa {\em et.~al.} \cite{ochoa:19} that a lower-energy state can be generated by overlapping two excited states via an ICM update. Therefore, one interesting addition to TCA could be trying to push the pool states further down in energy by performing ICM updates at the end of the cycle, provided that the graph density is low enough and the clusters do not percolate. 

\begin{acknowledgments}

We would like to thank A.~M{\"o}bius for useful discussions regarding various aspects of
the thermal cycling algorithm, and also providing code for the double-spin random moves (DSRM).
The authors would also like to acknowledge Jonathan Machta for critically reviewing the
manuscript. H.G.K.~would like to thanks Dr.~Pimple Popper for visualizations of energy
landscapes. We thank Texas A\&M University, NASA Ames Research Center, and Microsoft Quantum Group for providing access to computational resources. This work is supported in part by the Office of the Director of National Intelligence (ODNI), Intelligence Advanced Research Projects Activity (IARPA), via MIT Lincoln Laboratory Air Force Contract No.~FA8721-05-C-0002. SM also acknowledges the support from the Intelligence Advanced Research Projects Activity (IARPA) -- IARPA~IAA~1198 -- and the Defense Advanced Research Projects Agency (DARPA) -- IAA~8839, Annex~125 --. The views and conclusions contained herein are those of the authors and should not be interpreted as necessarily representing the official policies or endorsements, either expressed or implied, of ODNI, IARPA, DARPA or the U.S. Government. The U.S.  Government is authorized to reproduce and distribute reprints for Governmental purposes notwithstanding any copyright annotation thereon.

\end{acknowledgments}
\bibliographystyle{apsrev}
\bibliography{refs}

\begin{thebibliography}{63}
\expandafter\ifx\csname natexlab\endcsname\relax\def\natexlab#1{#1}\fi
\expandafter\ifx\csname bibnamefont\endcsname\relax
  \def\bibnamefont#1{#1}\fi
\expandafter\ifx\csname bibfnamefont\endcsname\relax
  \def\bibfnamefont#1{#1}\fi
\expandafter\ifx\csname citenamefont\endcsname\relax
  \def\citenamefont#1{#1}\fi
\expandafter\ifx\csname url\endcsname\relax
  \def\url#1{\texttt{#1}}\fi
\expandafter\ifx\csname urlprefix\endcsname\relax\def\urlprefix{URL }\fi
\providecommand{\bibinfo}[2]{#2}
\providecommand{\eprint}[2][]{\url{#2}}

\bibitem[{\citenamefont{Stanev et~al.}(2018)\citenamefont{Stanev, Oses, Kusne,
  Rodriguez, Curtarolo, and Takeuchi}}]{stanev:18}
\bibinfo{author}{\bibfnamefont{V.}~\bibnamefont{Stanev}},
  \bibinfo{author}{\bibfnamefont{C.}~\bibnamefont{Oses}},
  \bibinfo{author}{\bibfnamefont{A.~G.} \bibnamefont{Kusne}},
  \bibinfo{author}{\bibfnamefont{E.}~\bibnamefont{Rodriguez}},
  \bibinfo{author}{\bibfnamefont{S.}~\bibnamefont{Curtarolo}},
  \bibnamefont{and} \bibinfo{author}{\bibfnamefont{I.}~\bibnamefont{Takeuchi}},
  \bibinfo{journal}{npj Computational Materials} \textbf{\bibinfo{volume}{4}},
  \bibinfo{pages}{29} (\bibinfo{year}{2018}).

\bibitem[{\citenamefont{Bends{\o}e}(2009)}]{bendsoe:09}
\bibinfo{author}{\bibfnamefont{M.~P.} \bibnamefont{Bends{\o}e}},
  \emph{\bibinfo{title}{Topology optimizationTopology Optimization}}
  (\bibinfo{publisher}{Springer US}, \bibinfo{address}{Boston, MA},
  \bibinfo{year}{2009}), pp. \bibinfo{pages}{3928--3929}.

\bibitem[{\citenamefont{Keserű and Makara}(2006)}]{keseru:06}
\bibinfo{author}{\bibfnamefont{G.~M.} \bibnamefont{Keserű}} \bibnamefont{and}
  \bibinfo{author}{\bibfnamefont{G.~M.} \bibnamefont{Makara}},
  \bibinfo{journal}{Drug Discovery Today} \textbf{\bibinfo{volume}{11}},
  \bibinfo{pages}{741 } (\bibinfo{year}{2006}), ISSN \bibinfo{issn}{1359-6446}.

\bibitem[{\citenamefont{Shirazi et~al.}(2018)\citenamefont{Shirazi, Ceberio,
  and Lozano}}]{shirazi:18}
\bibinfo{author}{\bibfnamefont{A.}~\bibnamefont{Shirazi}},
  \bibinfo{author}{\bibfnamefont{J.}~\bibnamefont{Ceberio}}, \bibnamefont{and}
  \bibinfo{author}{\bibfnamefont{J.~A.} \bibnamefont{Lozano}},
  \bibinfo{journal}{Progress in Aerospace Sciences}
  \textbf{\bibinfo{volume}{102}}, \bibinfo{pages}{76 } (\bibinfo{year}{2018}),
  ISSN \bibinfo{issn}{0376-0421}.

\bibitem[{\citenamefont{Doering et~al.}(2016)\citenamefont{Doering, Juan,
  Kizys, Fito, and Calvet}}]{doering:16}
\bibinfo{author}{\bibfnamefont{J.}~\bibnamefont{Doering}},
  \bibinfo{author}{\bibfnamefont{A.~A.} \bibnamefont{Juan}},
  \bibinfo{author}{\bibfnamefont{R.}~\bibnamefont{Kizys}},
  \bibinfo{author}{\bibfnamefont{A.}~\bibnamefont{Fito}}, \bibnamefont{and}
  \bibinfo{author}{\bibfnamefont{L.}~\bibnamefont{Calvet}}, in
  \emph{\bibinfo{booktitle}{Modeling and Simulation in Engineering, Economics
  and Management}}, edited by
  \bibinfo{editor}{\bibfnamefont{R.}~\bibnamefont{Le{\'o}n}},
  \bibinfo{editor}{\bibfnamefont{M.~J.} \bibnamefont{Mu{\~{n}}oz-Torres}},
  \bibnamefont{and} \bibinfo{editor}{\bibfnamefont{J.~M.} \bibnamefont{Moneva}}
  (\bibinfo{publisher}{Springer International Publishing},
  \bibinfo{address}{Cham}, \bibinfo{year}{2016}), pp. \bibinfo{pages}{22--30}.

\bibitem[{\citenamefont{Guihaire and Hao}(2008)}]{guihaire:08}
\bibinfo{author}{\bibfnamefont{V.}~\bibnamefont{Guihaire}} \bibnamefont{and}
  \bibinfo{author}{\bibfnamefont{J.-K.} \bibnamefont{Hao}},
  \bibinfo{journal}{Transportation Research Part A: Policy and Practice}
  \textbf{\bibinfo{volume}{42}}, \bibinfo{pages}{1251 } (\bibinfo{year}{2008}),
  ISSN \bibinfo{issn}{0965-8564}.

\bibitem[{\citenamefont{Kinnunen and Li}(2010)}]{kinnunen:10}
\bibinfo{author}{\bibfnamefont{T.}~\bibnamefont{Kinnunen}} \bibnamefont{and}
  \bibinfo{author}{\bibfnamefont{H.}~\bibnamefont{Li}},
  \bibinfo{journal}{Speech Communication} \textbf{\bibinfo{volume}{52}},
  \bibinfo{pages}{12 } (\bibinfo{year}{2010}), ISSN \bibinfo{issn}{0167-6393}.

\bibitem[{\citenamefont{Karp}(1972)}]{karp:72}
\bibinfo{author}{\bibfnamefont{R.~M.} \bibnamefont{Karp}},
  \emph{\bibinfo{title}{{{Complexity of Computer Computations}}}}
  (\bibinfo{publisher}{New York: Plenum}, \bibinfo{year}{1972}),
  p.~\bibinfo{pages}{85}.

\bibitem[{\citenamefont{Cormen et~al.}(1990)\citenamefont{Cormen, Leiserson,
  and Rivest}}]{cormen:90}
\bibinfo{author}{\bibfnamefont{T.~H.} \bibnamefont{Cormen}},
  \bibinfo{author}{\bibfnamefont{T.~E.} \bibnamefont{Leiserson}},
  \bibnamefont{and} \bibinfo{author}{\bibfnamefont{R.~L.}
  \bibnamefont{Rivest}}, \emph{\bibinfo{title}{Introduction to Algorithms}}
  (\bibinfo{publisher}{MIT Press}, \bibinfo{address}{Cambridge, MA},
  \bibinfo{year}{1990}).

\bibitem[{\citenamefont{{Bendall} and {Margot}}(2006)}]{bendall:06}
\bibinfo{author}{\bibfnamefont{G.}~\bibnamefont{{Bendall}}} \bibnamefont{and}
  \bibinfo{author}{\bibfnamefont{F.}~\bibnamefont{{Margot}}},
  \bibinfo{journal}{Discrete Optimization} \textbf{\bibinfo{volume}{3}},
  \bibinfo{pages}{288 } (\bibinfo{year}{2006}).

\bibitem[{\citenamefont{Metropolis and Ulam}(1949)}]{metropolis:49}
\bibinfo{author}{\bibfnamefont{N.}~\bibnamefont{Metropolis}} \bibnamefont{and}
  \bibinfo{author}{\bibfnamefont{S.}~\bibnamefont{Ulam}}, \bibinfo{journal}{J.
  Am. Stat. Assoc.} \textbf{\bibinfo{volume}{44}}, \bibinfo{pages}{335}
  (\bibinfo{year}{1949}).

\bibitem[{\citenamefont{Kirkpatrick et~al.}(1983)\citenamefont{Kirkpatrick,
  {Gelatt, Jr.}, and Vecchi}}]{kirkpatrick:83}
\bibinfo{author}{\bibfnamefont{S.}~\bibnamefont{Kirkpatrick}},
  \bibinfo{author}{\bibfnamefont{C.~D.} \bibnamefont{{Gelatt, Jr.}}},
  \bibnamefont{and} \bibinfo{author}{\bibfnamefont{M.~P.}
  \bibnamefont{Vecchi}}, \bibinfo{journal}{Science}
  \textbf{\bibinfo{volume}{220}}, \bibinfo{pages}{671} (\bibinfo{year}{1983}).

\bibitem[{\citenamefont{Geman and Geman}(1984)}]{geman:84}
\bibinfo{author}{\bibfnamefont{S.}~\bibnamefont{Geman}} \bibnamefont{and}
  \bibinfo{author}{\bibfnamefont{D.}~\bibnamefont{Geman}},
  \bibinfo{journal}{IEEE Trans. Pattern. Analy. Mach. Intell.}
  \textbf{\bibinfo{volume}{PAMI-6}}, \bibinfo{pages}{721}
  (\bibinfo{year}{1984}).

\bibitem[{\citenamefont{{Trotter}}(1959)}]{trotter:59}
\bibinfo{author}{\bibfnamefont{H.~F.} \bibnamefont{{Trotter}}},
  \bibinfo{journal}{Proc. Amer. Math. Soc} \textbf{\bibinfo{volume}{10}},
  \bibinfo{pages}{545} (\bibinfo{year}{1959}).

\bibitem[{\citenamefont{Suzuki}(1971)}]{suzuki:71}
\bibinfo{author}{\bibfnamefont{M.}~\bibnamefont{Suzuki}},
  \bibinfo{journal}{Progress of Theoretical Physics}
  \textbf{\bibinfo{volume}{46}}, \bibinfo{pages}{1337} (\bibinfo{year}{1971}).

\bibitem[{\citenamefont{Suzuki}(1976)}]{suzuki:76}
\bibinfo{author}{\bibfnamefont{M.}~\bibnamefont{Suzuki}},
  \bibinfo{journal}{Progress of Theoretical Physics}
  \textbf{\bibinfo{volume}{56}}, \bibinfo{pages}{1454} (\bibinfo{year}{1976}).

\bibitem[{\citenamefont{Suzuki}(1993)}]{suzuki:93}
\bibinfo{author}{\bibfnamefont{M.}~\bibnamefont{Suzuki}},
  \emph{\bibinfo{title}{{Quantum Monte Carlo Methods in Condensed Matter
  Physics}}} (\bibinfo{publisher}{World Scientific},
  \bibinfo{address}{Singapore}, \bibinfo{year}{1993}).

\bibitem[{\citenamefont{{Landau} and {Binder}}(2000)}]{landau:00}
\bibinfo{author}{\bibfnamefont{D.~P.} \bibnamefont{{Landau}}} \bibnamefont{and}
  \bibinfo{author}{\bibfnamefont{K.}~\bibnamefont{{Binder}}},
  \emph{\bibinfo{title}{{A Guide to Monte Carlo Simulations in Statistical
  Physics}}} (\bibinfo{publisher}{Cambridge University Press},
  \bibinfo{year}{2000}).

\bibitem[{\citenamefont{{Troyer} et~al.}(2003)\citenamefont{{Troyer}, {Alet},
  {Trebst}, and {Wessel}}}]{troyer:03}
\bibinfo{author}{\bibfnamefont{M.}~\bibnamefont{{Troyer}}},
  \bibinfo{author}{\bibfnamefont{F.}~\bibnamefont{{Alet}}},
  \bibinfo{author}{\bibfnamefont{S.}~\bibnamefont{{Trebst}}}, \bibnamefont{and}
  \bibinfo{author}{\bibfnamefont{S.}~\bibnamefont{{Wessel}}}, in
  \emph{\bibinfo{booktitle}{AIP Conf. Proc. 690: The Monte Carlo Method in the
  Physical Sciences}} (\bibinfo{year}{2003}), pp. \bibinfo{pages}{156--169}.

\bibitem[{\citenamefont{Geyer}(1991)}]{geyer:91}
\bibinfo{author}{\bibfnamefont{C.}~\bibnamefont{Geyer}}, in
  \emph{\bibinfo{booktitle}{23rd Symposium on the Interface}}, edited by
  \bibinfo{editor}{\bibfnamefont{E.~M.} \bibnamefont{Keramidas}}
  (\bibinfo{publisher}{Interface Foundation}, \bibinfo{address}{Fairfax
  Station, VA}, \bibinfo{year}{1991}), p. \bibinfo{pages}{156}.

\bibitem[{\citenamefont{Hukushima and Nemoto}(1996)}]{hukushima:96}
\bibinfo{author}{\bibfnamefont{K.}~\bibnamefont{Hukushima}} \bibnamefont{and}
  \bibinfo{author}{\bibfnamefont{K.}~\bibnamefont{Nemoto}},
  \bibinfo{journal}{J. Phys. Soc. Jpn.} \textbf{\bibinfo{volume}{65}},
  \bibinfo{pages}{1604} (\bibinfo{year}{1996}).

\bibitem[{\citenamefont{Hukushima and Iba}(2003)}]{hukushima:03}
\bibinfo{author}{\bibfnamefont{K.}~\bibnamefont{Hukushima}} \bibnamefont{and}
  \bibinfo{author}{\bibfnamefont{Y.}~\bibnamefont{Iba}}, in
  \emph{\bibinfo{booktitle}{{The Monte Carlo method in the physical sciences:
  celebrating the 50th anniversary of the Metropolis algorithm}}}, edited by
  \bibinfo{editor}{\bibfnamefont{J.~E.} \bibnamefont{Gubernatis}}
  (\bibinfo{publisher}{AIP}, \bibinfo{address}{Los Alamos, New Mexico (USA)},
  \bibinfo{year}{2003}), vol. \bibinfo{volume}{690}, p. \bibinfo{pages}{200}.

\bibitem[{\citenamefont{Machta}(2010)}]{machta:10}
\bibinfo{author}{\bibfnamefont{J.}~\bibnamefont{Machta}},
  \bibinfo{journal}{Phys. Rev. E} \textbf{\bibinfo{volume}{82}},
  \bibinfo{pages}{026704} (\bibinfo{year}{2010}).

\bibitem[{\citenamefont{Wang et~al.}(2015)\citenamefont{Wang, Machta, and
  Katzgraber}}]{wang:15e}
\bibinfo{author}{\bibfnamefont{W.}~\bibnamefont{Wang}},
  \bibinfo{author}{\bibfnamefont{J.}~\bibnamefont{Machta}}, \bibnamefont{and}
  \bibinfo{author}{\bibfnamefont{H.~G.} \bibnamefont{Katzgraber}},
  \bibinfo{journal}{Phys. Rev. E} \textbf{\bibinfo{volume}{92}},
  \bibinfo{pages}{063307} (\bibinfo{year}{2015}).

\bibitem[{\citenamefont{Amey and Machta}(2018)}]{amey:18}
\bibinfo{author}{\bibfnamefont{C.}~\bibnamefont{Amey}} \bibnamefont{and}
  \bibinfo{author}{\bibfnamefont{J.}~\bibnamefont{Machta}},
  \bibinfo{journal}{Phys. Rev. E} \textbf{\bibinfo{volume}{97}},
  \bibinfo{pages}{033301} (\bibinfo{year}{2018}).

\bibitem[{\citenamefont{{Barzegar} et~al.}(2018)\citenamefont{{Barzegar},
  {Pattison}, {Wang}, and {Katzgraber}}}]{barzegar:18a}
\bibinfo{author}{\bibfnamefont{A.}~\bibnamefont{{Barzegar}}},
  \bibinfo{author}{\bibfnamefont{C.}~\bibnamefont{{Pattison}}},
  \bibinfo{author}{\bibfnamefont{W.}~\bibnamefont{{Wang}}}, \bibnamefont{and}
  \bibinfo{author}{\bibfnamefont{H.~G.} \bibnamefont{{Katzgraber}}},
  \bibinfo{journal}{Phys. Rev. E} \textbf{\bibinfo{volume}{98}},
  \bibinfo{pages}{053308} (\bibinfo{year}{2018}).

\bibitem[{\citenamefont{Brady}(1985)}]{brady:85}
\bibinfo{author}{\bibfnamefont{R.~M.} \bibnamefont{Brady}},
  \bibinfo{journal}{Nature} \textbf{\bibinfo{volume}{317}},
  \bibinfo{pages}{804} (\bibinfo{year}{1985}).

\bibitem[{\citenamefont{M{\"u}hlenbein
  et~al.}(1988)\citenamefont{M{\"u}hlenbein, Gorges-Schleuter, and
  Kr{\"a}mer}}]{muhlenbein:88}
\bibinfo{author}{\bibfnamefont{H.}~\bibnamefont{M{\"u}hlenbein}},
  \bibinfo{author}{\bibfnamefont{M.}~\bibnamefont{Gorges-Schleuter}},
  \bibnamefont{and}
  \bibinfo{author}{\bibfnamefont{O.}~\bibnamefont{Kr{\"a}mer}},
  \bibinfo{journal}{Parallel Computing} \textbf{\bibinfo{volume}{7}},
  \bibinfo{pages}{65 } (\bibinfo{year}{1988}), ISSN \bibinfo{issn}{0167-8191}.

\bibitem[{\citenamefont{Freisleben and Merz}(1996)}]{freisleben:96}
\bibinfo{author}{\bibfnamefont{B.}~\bibnamefont{Freisleben}} \bibnamefont{and}
  \bibinfo{author}{\bibfnamefont{P.}~\bibnamefont{Merz}}, in
  \emph{\bibinfo{booktitle}{Proceedings of 1996 {IEEE} International Conference
  on Evolutionary Computation, Nayoya University, Japan, May 20-22, 1996}},
  edited by \bibinfo{editor}{\bibfnamefont{T.}~\bibnamefont{Fukuda}}
  \bibnamefont{and} \bibinfo{editor}{\bibfnamefont{T.}~\bibnamefont{Furuhashi}}
  (\bibinfo{publisher}{{IEEE}}, \bibinfo{year}{1996}), pp.
  \bibinfo{pages}{616--621}.

\bibitem[{\citenamefont{Merz and Freisleben}(1998)}]{merz:98}
\bibinfo{author}{\bibfnamefont{P.}~\bibnamefont{Merz}} \bibnamefont{and}
  \bibinfo{author}{\bibfnamefont{B.}~\bibnamefont{Freisleben}},
  \emph{\bibinfo{title}{Lecture Notes in Computer Sci. 1498}}
  (\bibinfo{publisher}{Springer, Berlin}, \bibinfo{year}{1998}), p.
  \bibinfo{pages}{765}.

\bibitem[{\citenamefont{M\"obius et~al.}(1997)\citenamefont{M\"obius,
  Neklioudov, D\'{\i}az-S\'anchez, Hoffmann, Fachat, and
  Schreiber}}]{moebius:97}
\bibinfo{author}{\bibfnamefont{A.}~\bibnamefont{M\"obius}},
  \bibinfo{author}{\bibfnamefont{A.}~\bibnamefont{Neklioudov}},
  \bibinfo{author}{\bibfnamefont{A.}~\bibnamefont{D\'{\i}az-S\'anchez}},
  \bibinfo{author}{\bibfnamefont{K.~H.} \bibnamefont{Hoffmann}},
  \bibinfo{author}{\bibfnamefont{A.}~\bibnamefont{Fachat}}, \bibnamefont{and}
  \bibinfo{author}{\bibfnamefont{M.}~\bibnamefont{Schreiber}},
  \bibinfo{journal}{Phys. Rev. Lett.} \textbf{\bibinfo{volume}{79}},
  \bibinfo{pages}{4297} (\bibinfo{year}{1997}).

\bibitem[{\citenamefont{M\"obius et~al.}(2005)\citenamefont{M\"obius, Hoffmann,
  and Sch\"on}}]{moebius:05}
\bibinfo{author}{\bibfnamefont{A.}~\bibnamefont{M\"obius}},
  \bibinfo{author}{\bibfnamefont{K.~H.} \bibnamefont{Hoffmann}},
  \bibnamefont{and} \bibinfo{author}{\bibfnamefont{C.}~\bibnamefont{Sch\"on}},
  \emph{\bibinfo{title}{Optimization by Thermal Cycling}}
  (\bibinfo{publisher}{World Scientific}, \bibinfo{address}{Singapore;
  Hackensack, N.J}, \bibinfo{year}{2005}), p. \bibinfo{pages}{215}.

\bibitem[{\citenamefont{{Boixo} et~al.}(2014)\citenamefont{{Boixo},
  {R{\o}nnow}, {Isakov}, {Wang}, {Wecker}, {Lidar}, {Martinis}, and
  {Troyer}}}]{boixo:14}
\bibinfo{author}{\bibfnamefont{S.}~\bibnamefont{{Boixo}}},
  \bibinfo{author}{\bibfnamefont{T.~F.} \bibnamefont{{R{\o}nnow}}},
  \bibinfo{author}{\bibfnamefont{S.~V.} \bibnamefont{{Isakov}}},
  \bibinfo{author}{\bibfnamefont{Z.}~\bibnamefont{{Wang}}},
  \bibinfo{author}{\bibfnamefont{D.}~\bibnamefont{{Wecker}}},
  \bibinfo{author}{\bibfnamefont{D.~A.} \bibnamefont{{Lidar}}},
  \bibinfo{author}{\bibfnamefont{J.~M.} \bibnamefont{{Martinis}}},
  \bibnamefont{and} \bibinfo{author}{\bibfnamefont{M.}~\bibnamefont{{Troyer}}},
  \bibinfo{journal}{Nat. Phys.} \textbf{\bibinfo{volume}{10}},
  \bibinfo{pages}{218} (\bibinfo{year}{2014}).

\bibitem[{\citenamefont{{R{\o}nnow} et~al.}(2014)\citenamefont{{R{\o}nnow},
  {Wang}, {Job}, {Boixo}, {Isakov}, {Wecker}, {Martinis}, {Lidar}, and
  {Troyer}}}]{ronnow:14a}
\bibinfo{author}{\bibfnamefont{T.~F.} \bibnamefont{{R{\o}nnow}}},
  \bibinfo{author}{\bibfnamefont{Z.}~\bibnamefont{{Wang}}},
  \bibinfo{author}{\bibfnamefont{J.}~\bibnamefont{{Job}}},
  \bibinfo{author}{\bibfnamefont{S.}~\bibnamefont{{Boixo}}},
  \bibinfo{author}{\bibfnamefont{S.~V.} \bibnamefont{{Isakov}}},
  \bibinfo{author}{\bibfnamefont{D.}~\bibnamefont{{Wecker}}},
  \bibinfo{author}{\bibfnamefont{J.~M.} \bibnamefont{{Martinis}}},
  \bibinfo{author}{\bibfnamefont{D.~A.} \bibnamefont{{Lidar}}},
  \bibnamefont{and} \bibinfo{author}{\bibfnamefont{M.}~\bibnamefont{{Troyer}}},
  \bibinfo{journal}{Science} \textbf{\bibinfo{volume}{345}},
  \bibinfo{pages}{420} (\bibinfo{year}{2014}).

\bibitem[{\citenamefont{Mandr\`{a} and Katzgraber}(2018)}]{mandra:18}
\bibinfo{author}{\bibfnamefont{S.}~\bibnamefont{Mandr\`{a}}} \bibnamefont{and}
  \bibinfo{author}{\bibfnamefont{H.~G.} \bibnamefont{Katzgraber}},
  \bibinfo{journal}{Quantum Sci. Technol.} \textbf{\bibinfo{volume}{3}},
  \bibinfo{pages}{04LT01} (\bibinfo{year}{2018}).

\bibitem[{dwa()}]{dwave}
\bibinfo{note}{D-Wave Systems Inc. The D-Wave 2X Quantum Computer}.

\bibitem[{\citenamefont{Hen et~al.}(2015)\citenamefont{Hen, Job, Albash,
  R{\o}nnow, Troyer, and Lidar}}]{hen:15a}
\bibinfo{author}{\bibfnamefont{I.}~\bibnamefont{Hen}},
  \bibinfo{author}{\bibfnamefont{J.}~\bibnamefont{Job}},
  \bibinfo{author}{\bibfnamefont{T.}~\bibnamefont{Albash}},
  \bibinfo{author}{\bibfnamefont{T.~F.} \bibnamefont{R{\o}nnow}},
  \bibinfo{author}{\bibfnamefont{M.}~\bibnamefont{Troyer}}, \bibnamefont{and}
  \bibinfo{author}{\bibfnamefont{D.~A.} \bibnamefont{Lidar}},
  \bibinfo{journal}{Phys. Rev. A} \textbf{\bibinfo{volume}{92}},
  \bibinfo{pages}{042325} (\bibinfo{year}{2015}).

\bibitem[{\citenamefont{{King} et~al.}(2019)\citenamefont{{King}, {Yarkoni},
  {Raymond}, {Ozfidan}, {King}, {Nevisi}, {Hilton}, and {McGeoch}}}]{king:19}
\bibinfo{author}{\bibfnamefont{J.}~\bibnamefont{{King}}},
  \bibinfo{author}{\bibfnamefont{S.}~\bibnamefont{{Yarkoni}}},
  \bibinfo{author}{\bibfnamefont{J.}~\bibnamefont{{Raymond}}},
  \bibinfo{author}{\bibfnamefont{I.}~\bibnamefont{{Ozfidan}}},
  \bibinfo{author}{\bibfnamefont{A.~D.} \bibnamefont{{King}}},
  \bibinfo{author}{\bibfnamefont{M.~M.} \bibnamefont{{Nevisi}}},
  \bibinfo{author}{\bibfnamefont{J.~P.} \bibnamefont{{Hilton}}},
  \bibnamefont{and} \bibinfo{author}{\bibfnamefont{C.~C.}
  \bibnamefont{{McGeoch}}}, \bibinfo{journal}{Journal of the Physical Society
  of Japan} \textbf{\bibinfo{volume}{88}}, \bibinfo{pages}{061007}
  (\bibinfo{year}{2019}).

\bibitem[{\citenamefont{Bunyk et~al.}(2014)\citenamefont{Bunyk, Hoskinson,
  Johnson, Tolkacheva, Altomare, Berkley, Harris, Hilton, Lanting, and
  Whittaker}}]{bunyk:14}
\bibinfo{author}{\bibfnamefont{P.}~\bibnamefont{Bunyk}},
  \bibinfo{author}{\bibfnamefont{E.}~\bibnamefont{Hoskinson}},
  \bibinfo{author}{\bibfnamefont{M.~W.} \bibnamefont{Johnson}},
  \bibinfo{author}{\bibfnamefont{E.}~\bibnamefont{Tolkacheva}},
  \bibinfo{author}{\bibfnamefont{F.}~\bibnamefont{Altomare}},
  \bibinfo{author}{\bibfnamefont{A.~J.} \bibnamefont{Berkley}},
  \bibinfo{author}{\bibfnamefont{R.}~\bibnamefont{Harris}},
  \bibinfo{author}{\bibfnamefont{J.~P.} \bibnamefont{Hilton}},
  \bibinfo{author}{\bibfnamefont{T.}~\bibnamefont{Lanting}}, \bibnamefont{and}
  \bibinfo{author}{\bibfnamefont{J.}~\bibnamefont{Whittaker}},
  \bibinfo{journal}{IEEE Trans. Appl. Supercond.}
  \textbf{\bibinfo{volume}{24}}, \bibinfo{pages}{1} (\bibinfo{year}{2014}).

\bibitem[{\citenamefont{King et~al.}(2015)\citenamefont{King, Lanting, and
  Harris}}]{king:15}
\bibinfo{author}{\bibfnamefont{A.~D.} \bibnamefont{King}},
  \bibinfo{author}{\bibfnamefont{T.}~\bibnamefont{Lanting}}, \bibnamefont{and}
  \bibinfo{author}{\bibfnamefont{R.}~\bibnamefont{Harris}},
  \bibinfo{journal}{arXiv:1502.02098 [quant-ph]}  (\bibinfo{year}{2015}).

\bibitem[{\citenamefont{{King} et~al.}(2017)\citenamefont{{King}, {Yarkoni},
  {Raymond}, {Ozfidan}, {King}, {Nevisi}, {Hilton}, and {McGeoch}}}]{king:17}
\bibinfo{author}{\bibfnamefont{J.}~\bibnamefont{{King}}},
  \bibinfo{author}{\bibfnamefont{S.}~\bibnamefont{{Yarkoni}}},
  \bibinfo{author}{\bibfnamefont{J.}~\bibnamefont{{Raymond}}},
  \bibinfo{author}{\bibfnamefont{I.}~\bibnamefont{{Ozfidan}}},
  \bibinfo{author}{\bibfnamefont{A.~D.} \bibnamefont{{King}}},
  \bibinfo{author}{\bibfnamefont{M.~M.} \bibnamefont{{Nevisi}}},
  \bibinfo{author}{\bibfnamefont{J.~P.} \bibnamefont{{Hilton}}},
  \bibnamefont{and} \bibinfo{author}{\bibfnamefont{C.~C.}
  \bibnamefont{{McGeoch}}}, \bibinfo{journal}{arXiv:1701.04579 [quant-ph]}
  (\bibinfo{year}{2017}).

\bibitem[{\citenamefont{Lin and Kernighan}(1973)}]{lin:73}
\bibinfo{author}{\bibfnamefont{S.}~\bibnamefont{Lin}} \bibnamefont{and}
  \bibinfo{author}{\bibfnamefont{B.~W.} \bibnamefont{Kernighan}},
  \bibinfo{journal}{Operations Research} \textbf{\bibinfo{volume}{21}},
  \bibinfo{pages}{498} (\bibinfo{year}{1973}), ISSN \bibinfo{issn}{0030364X,
  15265463}.

\bibitem[{\citenamefont{Helsgaun}(2000)}]{helsgaun:00}
\bibinfo{author}{\bibfnamefont{K.}~\bibnamefont{Helsgaun}},
  \bibinfo{journal}{European Journal of Operational Research}
  \textbf{\bibinfo{volume}{126}}, \bibinfo{pages}{106 } (\bibinfo{year}{2000}),
  ISSN \bibinfo{issn}{0377-2217}.

\bibitem[{\citenamefont{Flood}(1956)}]{flood:56}
\bibinfo{author}{\bibfnamefont{M.~M.} \bibnamefont{Flood}},
  \bibinfo{journal}{Operations Research} \textbf{\bibinfo{volume}{4}},
  \bibinfo{pages}{61} (\bibinfo{year}{1956}).

\bibitem[{\citenamefont{Croes}(1958)}]{croes:58}
\bibinfo{author}{\bibfnamefont{G.~A.} \bibnamefont{Croes}},
  \bibinfo{journal}{Operations Research} \textbf{\bibinfo{volume}{6}},
  \bibinfo{pages}{791} (\bibinfo{year}{1958}), ISSN \bibinfo{issn}{0030364X,
  15265463}.

\bibitem[{\citenamefont{Lin}(1965)}]{lin:65}
\bibinfo{author}{\bibfnamefont{S.}~\bibnamefont{Lin}}, \bibinfo{journal}{Bell
  System Technical Journal} \textbf{\bibinfo{volume}{44}},
  \bibinfo{pages}{2245} (\bibinfo{year}{1965}).

\bibitem[{\citenamefont{Glover}(1989)}]{glover:89}
\bibinfo{author}{\bibfnamefont{F.}~\bibnamefont{Glover}},
  \bibinfo{journal}{ORSA Journal on Computing} \textbf{\bibinfo{volume}{1}},
  \bibinfo{pages}{190} (\bibinfo{year}{1989}).

\bibitem[{\citenamefont{Glover}(1990)}]{glover:90}
\bibinfo{author}{\bibfnamefont{F.}~\bibnamefont{Glover}},
  \bibinfo{journal}{ORSA Journal on Computing} \textbf{\bibinfo{volume}{2}},
  \bibinfo{pages}{4} (\bibinfo{year}{1990}).

\bibitem[{\citenamefont{Hamze et~al.}(2011)\citenamefont{Hamze, Wang, and
  de~Freitas}}]{hamze:11}
\bibinfo{author}{\bibfnamefont{F.}~\bibnamefont{Hamze}},
  \bibinfo{author}{\bibfnamefont{Z.}~\bibnamefont{Wang}}, \bibnamefont{and}
  \bibinfo{author}{\bibfnamefont{N.}~\bibnamefont{de~Freitas}},
  \bibinfo{journal}{arXiv:1111.5379 [stat.CO]}  (\bibinfo{year}{2011}).

\bibitem[{\citenamefont{Schneider et~al.}(1996)\citenamefont{Schneider,
  Froschhammer, Morgenstern, Husslein, and Singer}}]{schneider:96}
\bibinfo{author}{\bibfnamefont{J.}~\bibnamefont{Schneider}},
  \bibinfo{author}{\bibfnamefont{C.}~\bibnamefont{Froschhammer}},
  \bibinfo{author}{\bibfnamefont{I.}~\bibnamefont{Morgenstern}},
  \bibinfo{author}{\bibfnamefont{T.}~\bibnamefont{Husslein}}, \bibnamefont{and}
  \bibinfo{author}{\bibfnamefont{J.~M.} \bibnamefont{Singer}},
  \bibinfo{journal}{Computer Physics Communications}
  \textbf{\bibinfo{volume}{96}}, \bibinfo{pages}{173 } (\bibinfo{year}{1996}).

\bibitem[{\citenamefont{Zhang}(2004)}]{zhang:04}
\bibinfo{author}{\bibfnamefont{W.}~\bibnamefont{Zhang}},
  \bibinfo{journal}{Artif. Intell.} \textbf{\bibinfo{volume}{158}},
  \bibinfo{pages}{1} (\bibinfo{year}{2004}).

\bibitem[{\citenamefont{Wang et~al.}(2011)\citenamefont{Wang, L{\"u}, Glover,
  and Hao}}]{wang:11b}
\bibinfo{author}{\bibfnamefont{Y.}~\bibnamefont{Wang}},
  \bibinfo{author}{\bibfnamefont{Z.}~\bibnamefont{L{\"u}}},
  \bibinfo{author}{\bibfnamefont{F.}~\bibnamefont{Glover}}, \bibnamefont{and}
  \bibinfo{author}{\bibfnamefont{J.-K.} \bibnamefont{Hao}}, in
  \emph{\bibinfo{booktitle}{Evolutionary Computation in Combinatorial
  Optimization}}, edited by
  \bibinfo{editor}{\bibfnamefont{P.}~\bibnamefont{Merz}} \bibnamefont{and}
  \bibinfo{editor}{\bibfnamefont{J.-K.} \bibnamefont{Hao}}
  (\bibinfo{publisher}{Springer Berlin Heidelberg}, \bibinfo{address}{Berlin,
  Heidelberg}, \bibinfo{year}{2011}), pp. \bibinfo{pages}{72--83}.

\bibitem[{\citenamefont{Wang et~al.}(2013)\citenamefont{Wang, L{\"u}, Glover,
  and Hao}}]{wang:13d}
\bibinfo{author}{\bibfnamefont{Y.}~\bibnamefont{Wang}},
  \bibinfo{author}{\bibfnamefont{Z.}~\bibnamefont{L{\"u}}},
  \bibinfo{author}{\bibfnamefont{F.}~\bibnamefont{Glover}}, \bibnamefont{and}
  \bibinfo{author}{\bibfnamefont{J.-K.} \bibnamefont{Hao}},
  \bibinfo{journal}{Journal of Heuristics} \textbf{\bibinfo{volume}{19}},
  \bibinfo{pages}{679} (\bibinfo{year}{2013}).

\bibitem[{\citenamefont{Chardaire et~al.}(1995)\citenamefont{Chardaire, Lutton,
  and Sutter}}]{chardaire:95}
\bibinfo{author}{\bibfnamefont{P.}~\bibnamefont{Chardaire}},
  \bibinfo{author}{\bibfnamefont{J.~L.} \bibnamefont{Lutton}},
  \bibnamefont{and} \bibinfo{author}{\bibfnamefont{A.}~\bibnamefont{Sutter}},
  \bibinfo{journal}{European Journal of Operational Research}
  \textbf{\bibinfo{volume}{86}}, \bibinfo{pages}{565 } (\bibinfo{year}{1995}),
  ISSN \bibinfo{issn}{0377-2217}.

\bibitem[{\citenamefont{Karimi et~al.}(2017)\citenamefont{Karimi, Rosenberg,
  and Katzgraber}}]{karimi:17a}
\bibinfo{author}{\bibfnamefont{H.}~\bibnamefont{Karimi}},
  \bibinfo{author}{\bibfnamefont{G.}~\bibnamefont{Rosenberg}},
  \bibnamefont{and} \bibinfo{author}{\bibfnamefont{H.~G.}
  \bibnamefont{Katzgraber}}, \bibinfo{journal}{Phys. Rev. E}
  \textbf{\bibinfo{volume}{96}}, \bibinfo{pages}{043312}
  (\bibinfo{year}{2017}).

\bibitem[{\citenamefont{Wolff}(1989)}]{wolff:89}
\bibinfo{author}{\bibfnamefont{U.}~\bibnamefont{Wolff}},
  \bibinfo{journal}{Phys. Rev. Lett.} \textbf{\bibinfo{volume}{62}},
  \bibinfo{pages}{361} (\bibinfo{year}{1989}).

\bibitem[{\citenamefont{{Zhu} et~al.}(2015)\citenamefont{{Zhu}, {Ochoa}, and
  {Katzgraber}}}]{zhu:15b}
\bibinfo{author}{\bibfnamefont{Z.}~\bibnamefont{{Zhu}}},
  \bibinfo{author}{\bibfnamefont{A.~J.} \bibnamefont{{Ochoa}}},
  \bibnamefont{and} \bibinfo{author}{\bibfnamefont{H.~G.}
  \bibnamefont{{Katzgraber}}}, \bibinfo{journal}{Phys. Rev. Lett.}
  \textbf{\bibinfo{volume}{115}}, \bibinfo{pages}{077201}
  (\bibinfo{year}{2015}).

\bibitem[{\citenamefont{Finnila et~al.}(1994)\citenamefont{Finnila, Gomez,
  Sebenik, Stenson, and Doll}}]{finnila:94}
\bibinfo{author}{\bibfnamefont{A.~B.} \bibnamefont{Finnila}},
  \bibinfo{author}{\bibfnamefont{M.~A.} \bibnamefont{Gomez}},
  \bibinfo{author}{\bibfnamefont{C.}~\bibnamefont{Sebenik}},
  \bibinfo{author}{\bibfnamefont{C.}~\bibnamefont{Stenson}}, \bibnamefont{and}
  \bibinfo{author}{\bibfnamefont{J.~D.} \bibnamefont{Doll}},
  \bibinfo{journal}{Chem. Phys. Lett.} \textbf{\bibinfo{volume}{219}},
  \bibinfo{pages}{343} (\bibinfo{year}{1994}).

\bibitem[{\citenamefont{Kadowaki and Nishimori}(1998)}]{kadowaki:98}
\bibinfo{author}{\bibfnamefont{T.}~\bibnamefont{Kadowaki}} \bibnamefont{and}
  \bibinfo{author}{\bibfnamefont{H.}~\bibnamefont{Nishimori}},
  \bibinfo{journal}{Phys. Rev. E} \textbf{\bibinfo{volume}{58}},
  \bibinfo{pages}{5355} (\bibinfo{year}{1998}).

\bibitem[{\citenamefont{Born and Fock}(1928)}]{born:28}
\bibinfo{author}{\bibfnamefont{M.}~\bibnamefont{Born}} \bibnamefont{and}
  \bibinfo{author}{\bibfnamefont{V.}~\bibnamefont{Fock}},
  \bibinfo{journal}{Zeitschrift f{\"u}r Physik} \textbf{\bibinfo{volume}{51}},
  \bibinfo{pages}{165} (\bibinfo{year}{1928}).

\bibitem[{\citenamefont{M\"obius et~al.}(1999)\citenamefont{M\"obius,
  Freisleben, Merz, and Schreiber}}]{moebius:99a}
\bibinfo{author}{\bibfnamefont{A.}~\bibnamefont{M\"obius}},
  \bibinfo{author}{\bibfnamefont{B.}~\bibnamefont{Freisleben}},
  \bibinfo{author}{\bibfnamefont{P.}~\bibnamefont{Merz}}, \bibnamefont{and}
  \bibinfo{author}{\bibfnamefont{M.}~\bibnamefont{Schreiber}},
  \bibinfo{journal}{Phys. Rev. E} \textbf{\bibinfo{volume}{59}},
  \bibinfo{pages}{4667} (\bibinfo{year}{1999}).

\bibitem[{\citenamefont{M{\"o}bius et~al.}(1999)\citenamefont{M{\"o}bius,
  Diaz-Sanchez, Freisleben, Schreiber, Fachat, Hoffmann, Merz, and
  Neklioudov}}]{moebius:99b}
\bibinfo{author}{\bibfnamefont{A.}~\bibnamefont{M{\"o}bius}},
  \bibinfo{author}{\bibfnamefont{A.}~\bibnamefont{Diaz-Sanchez}},
  \bibinfo{author}{\bibfnamefont{B.}~\bibnamefont{Freisleben}},
  \bibinfo{author}{\bibfnamefont{M.}~\bibnamefont{Schreiber}},
  \bibinfo{author}{\bibfnamefont{A.}~\bibnamefont{Fachat}},
  \bibinfo{author}{\bibfnamefont{K.}~\bibnamefont{Hoffmann}},
  \bibinfo{author}{\bibfnamefont{P.}~\bibnamefont{Merz}}, \bibnamefont{and}
  \bibinfo{author}{\bibfnamefont{A.}~\bibnamefont{Neklioudov}},
  \bibinfo{journal}{Computer Physics Communications}
  \textbf{\bibinfo{volume}{121-122}}, \bibinfo{pages}{34 }
  (\bibinfo{year}{1999}), ISSN \bibinfo{issn}{0010-4655},
  \bibinfo{note}{proceedings of the Europhysics Conference on Computational
  Physics CCP 1998}.

\bibitem[{\citenamefont{Ochoa et~al.}(2019)\citenamefont{Ochoa, Jacob,
  Mandr\`a, and Katzgraber}}]{ochoa:19}
\bibinfo{author}{\bibfnamefont{A.~J.} \bibnamefont{Ochoa}},
  \bibinfo{author}{\bibfnamefont{D.~C.} \bibnamefont{Jacob}},
  \bibinfo{author}{\bibfnamefont{S.}~\bibnamefont{Mandr\`a}}, \bibnamefont{and}
  \bibinfo{author}{\bibfnamefont{H.~G.} \bibnamefont{Katzgraber}},
  \bibinfo{journal}{Phys. Rev. E} \textbf{\bibinfo{volume}{99}},
  \bibinfo{pages}{043306} (\bibinfo{year}{2019}).

\end{thebibliography}

\end{document}